\documentclass[useAMS,usenatbib]{mn2e}

\usepackage{psfig, epsf, epsfig}

\title[Origin of the Galactic bulge]{Formation of the Galactic bulge
from a two-component stellar disk: Explaining cylindrical rotation and
vertical metallicity gradient}
\author[K. Bekki and   T. Tsujimoto]
{Kenji Bekki${}^1$\thanks{E-mail:
bekki@cyllene.uwa.edu.au}
and Takuji  Tsujimoto${}^2$ \\
${}^1$ICRAR M468
The University of Western Australia
35 Stirling Hwy, Crawley
Western Australia, 6009 \\
${}^2$
National Astronomical Observatory, Mitaka-shi, Tokyo 181-8588, Japan \\}

\begin{document}

\date{Accepted, Received 2005 February 20; in original form }

\pagerange{\pageref{firstpage}--\pageref{lastpage}} \pubyear{2005}

\maketitle

\label{firstpage}

\begin{abstract}

Recent observational studies have revealed that the Galactic bulge 
has cylindrical rotation and a steeper vertical metallicity gradient.
We adopt two representative models for the bulge formation
and thereby investigate whether the two models can explain both the observed
cylindrical rotation and vertical metallicity gradient in a self-consistent manner.
One is the ``pure disk scenario'' (PDS) in which the bulge is 
formed from a pure thin  stellar disk through spontaneous bar
instability. The other
is the ``two-component disk scenario'' (TCDS) in which the bulge
is formed from a disk composed of thin and thick  disks through bar instability.
Our numerical simulations show that although PDS can reproduce the cylindrical rotation,
it shows a rather flatter vertical metallicity gradient that is inconsistent with
observations. The derived flatter metallicity gradient is due to the vertical mixing 
of stars with different initial metallicities by
the stellar bar.  This result implies that the bulge can not be simply formed
from  a pure thin stellar disk.
On the other hand,
the bulge formed from the two-component disk in TCDS
can explain both the observed cylindrical rotation  and  vertical metallicity
gradient of the Galactic bulge reasonably well.
In TCDS,
more metal-poor stars at higher $|z|$ (vertical distance) which originate from
the already dynamically hotter thick disk can not be strongly influenced by vertical
mixing of the bar  so that they can stay
in situ for longer timescales  and thus keep the lower metallicity at higher $|z|$.
Consequently, the vertical metallicity gradient of the bulge 
composed of initially thin and thick disk stars
can not be so flattened, even if the gradient of the thin disk 
can be flattened significantly by the bar in TCDS.
We therefore suggest that a significant fraction  of the present
Galactic bulge is composed of stars initially in the inner part of the
thick disk and thus that these bulge stars and the thick disk
have a common origin.
We also suggest that the Galaxy might well have experienced some merger 
events that could dynamically heat up its inner regions until $\sim 10$ Gyr ago.
\end{abstract}


\section{Introduction}

The Galaxy is observed to have a ``boxy'' bulge in near-infrared images
(Dwek et al. 1995) and the nature and the  origin of the inner triaxial shape
of the Galaxy (i.e., bar/bulge)
has been extensively
discussed by several authors (e.g., Babusiaux \& Gilmore 2005;
Rattenbury et al. 2007). 
Recent spectroscopic observations on stellar abundances and kinematics
of the Galactic bulge have provided new clues to the origin of the triaxial bulge
(e.g., Mel\'endez et al. 2008; Zoccali et al. 2008;
Babusiaux et al. 2010).
Mel\'endez et al. (2008) found no/little differences in
chemical abundances of stars between the Galactic bulge and the thick disk
and accordingly suggested a similar chemical evolution history between
the two components of the Galaxy (see also Ryde et al. 2010,  Alves-Brito
et al. 2010). Zocalli et al. (2008) found a  steep metallicity gradient
along the minor axis of the bulge
and suggested that the presence  of the vertical metallicity gradient
is consistent with dissipative processes of the bulge formation.
Babusiaux et al. (2010) revealed  that (i) the bulge has two distinct
populations and (i) the  metal-rich population
has bar-like kinematics whereas the metal-poor one has kinematics
similar to those of an  old spheroid or a thick disk
(see also de Propris et al. 2011).

Combes \& Sanders (1981) first pointed out that galactic bars can be
identified as boxy/peanut bulges in the edge-on view.
Dynamical models in which the Galactic bulge is formed from
the thin disk through spontaneous bar instability have been successfully 
constructed (e.g., Fux  1997) and used to discuss the bulge formation
in the context of PDS.
Although this PDS can explain consistently the observed cylindrical
rotation and velocity dispersion profiles of stars in the Galactic bulge
(e.g., Howard et al. 2009; Shen et al. 2010),  
recent observational studies have suggested that
the observed vertical metallicity gradient of the bulge can not be so simply
explained by the PDS (e.g., Minniti et al. 1995;
Zoccali et al. 2008; Babusiaux et al. 2010): evolution of the vertical
metallicity gradient  however has not been
extensively investigated by theoretical studies.

The purpose of this paper is thus to adopt
PDS and thereby to  investigate
the evolution of the vertical metallicity gradient of the Galactic bulge
formed from a pure thin stellar disk through bar instability
based on collisionless N-body numerical
simulations.
We also adopt TCDS and thereby investigate whether the simulated bulge can
explain both the observed cylindrical rotation and  
the vertical metallicity gradient in a self-consistent manner.
Given that the Galaxy is observed to have the thick disk 
as well as the thin one (e.g., Gilmore \& Reid 1983),
it is natural to adopt a two-component disk model for discussing the
origin of the bulge. 
%Previous simulations of major mergers 
%(that might well form
%galactic bulges)
%have shown radial/vertical metallicity gradients (e.g. Bekki \& Shioya 1999)
%but not yet reproduced 
%cylindrical rotation yet.
 
\section{The models}

\subsection{PDS}

The Galaxy  in PDS  is modeled as
a bulge-less disk galaxy with the total mass $M_{\rm d}$
and size $R_{\rm d}$ embedded in a massive dark matter halo.
The total  mass and the virial radius  of the dark matter halo  of  the disk
are denoted as $M_{\rm dm}$ ($=f_{\rm dm} M_{\rm d}$, 
where $f_{\rm dm}$ is the mass-ratio of dark matter to disk)
and $r_{\rm vir}$, respectively.
We adopted an NFW halo density distribution (Navarro, Frenk \& White 1996)
suggested from CDM simulations:
\begin{equation}
{\rho}(r)=\frac{\rho_{0}}{(r/r_{\rm s})(1+r/r_{\rm s})^2},
\end{equation}
where  $r$, $\rho_{0}$, and $r_{\rm s}$ are
the spherical radius,  the characteristic  density of a dark halo,  and the
scale
length of the halo, respectively.
The $c$ parameter ($=r_{\rm vir}/r_{\rm s}$)
is set to be 12 in the present study.

The stellar component of the disk  is assumed to have an exponential
profile with a radial scale length ($a$) 
and a vertical scale  height  ($h$).
In addition to the rotational velocity made by the gravitational
field of disk and halo component, the initial radial and azimuthal velocity
dispersion are given to the disk component according
to the epicyclic theory with Toomre's parameter (Binney \& Tremaine 1987) $Q$ = 1.5.
The vertical velocity dispersion at a given radius
is set to be 0.5 times as large as
the radial velocity dispersion at that point.
Although we have investigated different models with different parameters,
we describe the results of the model ``PDS1'' with $f_{\rm dm}=15.7$, 
$r_{\rm vir}=9R_{\rm d}$,
$M_{\rm d}=4 \times 10^{10} M_{\odot}$,  $R_{\rm d}=17.5$ kpc,
$a=1.75$ kpc, and $h=0.35$ kpc
in the present study, because this model clearly shows a bar (=boxy bulge)
with the cylindrical rotation consistent with  the observed one.

\subsection{TCDS}

Since our recent paper (Bekki \& Tsujimoto 2011) has already
described the methods and techniques
to construct two-component stellar disk models  in detail,
we briefly describe the  models here.
The Galaxy consists of the following two stellar components in TCDS.
One is a thick disk  
(or a  very flattened spheroid supported largely by rotation)
that was formed
as the first thin stellar disk (FD)
and later dynamically heated up by merger events.
The other is a thin disk (or second disk, SD) that was later formed via gas accretion
from the Galactic halo after the formation
of the thick disk. 
These dynamically hot and cold
stellar disks can be transformed into a bar (boxy bulge) through bar instability
in TCDS. The initial structural and kinematical properties of 
the stellar disk and the dark matter halo in FD 
and SD are essentially the same as those of the pure stellar disk adopted for PDS.  
The mass ($M_{\rm d, 1}$), size ($R_{\rm d, 1}$), 
scale-length ($a_1$), and scale-hight ($h_1$) of the initial 
exponential disk and the mass ($M_{\rm dm, 1}$), 
virial radius ($r_{\rm vir, 1}$), 
and $c$-parameter ($c_1$) of the dark matter halo in FD  are 
$8 \times 10^9 {\rm M}_{\odot}$, 17.5 kpc, 
1.75 kpc, 350 pc, $6.3 \times 10^{11} {\rm M}_{\odot}$,
210 kpc, and 12, respectively.

The initial disk merges with a dwarf disk galaxy so that a thick disk can be formed
within  $\sim 40$ dynamical timescale ($t_{\rm dyn}$)  of the initial disk
(Bekki \& Tsujimoto 2011). The initial disk and the dwarf are assumed to have
self-similar structural and kinematical properties of stellar disks and dark
matter halos.
The disk size of the dwarf ($R_{\rm d, dw}$) is assumed to depend on
the mass of the disk  ($M_{\rm d, dw}$) such as
$R_{\rm d, dw} \propto M_{\rm d,dw}^{0.5}$, which corresponds to
the Freeman's law (Freeman 1970). Therefore, the mass-ratio ($m_{2}$)
of the dwarf to  FD  can determine
the size of the stellar disk for the dwarf and thus the structural and
kinematical  properties of the dark matter halo and the stellar disk.
The stellar remnant of a minor merger with $m_2=0.2$, orbital
inclination angles of $30^{\circ}$,  orbital eccentricities of 0.5,
and pericenter distances of $R_{\rm d,1}$ (=17.5 kpc) are used as
the thick disk in the present study. After the thick disk
formation,  SD is assume to grow slowly for  $20 t_{\rm dyn}$
while the thick disk  and the dark matter halo can dynamically response to
the growing SD. The mass ($M_{\rm d, 2}$), 
size ($R_{\rm d, 2}$), scale-length ($a_2$), and scale-hight ($h_2$)
of the exponential stellar disk (SD) are $4 \times 10^{10} M_{\odot}$,
17.5 kpc, 3.5 kpc, and 350 pc, respectively.
The two-component disk constructed above is referred to as the ``TCDS1'' model.

\subsection{Radial and vertical metallicity gradients}

We adopt the same model for the radial metallicity gradient of a stellar disk
in PDS and TCDS
as that used in Bekki \& Tsujimoto (2011).
We consider that the metallicity gradient
is different between inner ($R<R_{\rm b}$)  and outer ($R \ge R_{\rm b}$)
regions of FD and SD, where
$R_{\rm b}$ is set to be 2 kpc corresponding to the size of the bulge.
We allocate metallicity to each disk star in the outer disk
($R \ge R_{\rm b}$)  according to its initial position:
the metallicity of the star 
at $r$ = $R$ (kpc)
is given as:
\begin{equation}
{\rm [m/H]}_{\rm r=R} = {\rm [m/H]}_{\rm d, r=0} + {\alpha}_{\rm d} \times {\rm R}. \;
\end{equation}
The constant  ${\rm [m/H]}_{\rm d, r=0}$ is determined such that the metallicity
at the solar radius ($R=R_{\odot}$ corresponding to 8.5 kpc) is consistent
with the observed one for a given
${\alpha}_{\rm d}$.
We adopt the observed  value  of ${\alpha}_{\rm d} \sim -0.04$
(e.g., Andrievsky et al. 2004).
In PDS, ${\rm [m/H]}_{\rm d, r=0}$ is set to be 0.14 
for ${\alpha}_{\rm d} \sim -0.04$ so that the mean metallicity
at $R=R_{\odot}$ can be $-0.2$.
In TCDS, ${\rm [m/H]}_{\rm d, r=0}$ is set to be $-0.36$ (0.14) for FD (SD)
for ${\alpha}_{\rm d} \sim -0.04$ so that the mean metallicity
at $R=R_{\odot}$ can be $-0.7$ ($-0.2$). 
The dwarf has ${\alpha}_{\rm d}=-0.04$ and
${\rm [m/H]}_{\rm d, r=0}=-0.54$ so that
the metallicity of the dwarf ($-0.18$ dex lower than FD)
is consistent with  the luminosity-metallicity relation
of galaxies ($M$ $\propto$ $Z^{0.4}$; Mould 1984).

We assign the metallicity of a star at $r$ = $R$
in the inner disk ($R<R_{\rm b}$),
where $r$ ($R$) is the projected distance (in units of kpc) of the star
from the center of the disk, to be:
\begin{equation}
{\rm [m/H]}_{\rm b, r=R} = {\rm [m/H]}_{\rm b, r=0} + {\alpha}_{\rm b} \times {\rm R}.
\;
\end{equation}
We consider  that the slope of the metallicity gradient is a free parameter and
investigate different models with different ${\alpha}_{\rm b}$.
The central metallicity of the disk is determined such that
the metallicity at $R=R_{\rm b}$ is consistent with the one derived from
the above equation (2).
For example, if we adopt 
${\alpha}_{\rm b} = -0.4$ for FD in TCDS,
then ${\rm [m/H]}_{\rm b, r=0} = 0.36$ for ${\alpha}_{\rm d} = -0.04$
and ${\rm [m/H]}_{\rm d, r=0} = -0.36$.
The inner radial metallicity gradients 
for initial disks in PDS and TCDS
are  denoted as ${\alpha}_{\rm b, pds}$ and ${\alpha}_{\rm b, tcds}$, respectively.

The initial metallicity
of a star with $r=R$ and  a vertical distance ($|z|$) from the $x$-$y$ 
(equatorial) plane  of the Galaxy
depends also on $|z|$ and 
it is assigned to be as follows 
\begin{equation}
{\rm [m/H]}_{\rm r=R, |z|} = {\rm [m/H]}_{|z|=0} + {\alpha}_{\rm v} \times |z|,
\end{equation}
where ${\rm [m/H]}_{\rm r=R, |z|=0}$ is a metallicity at $r=R$ and $|z|=0$.
Although ${\alpha}_{\rm v} \approx -0.4$ could be  reasonable (Frogel et al. 2000),
we investigate models with different  ${\alpha}_{\rm v}$.
The slope  ${\alpha}_{\rm v}$ is denoted as
${\alpha}_{\rm v, pds}$ and ${\alpha}_{\rm v, tcds}$ for PDS and TCDS,
respectively.

Using  the latest version of GRAPE
(GRavity PipE, GRAPE-DR) -- which is the special-purpose
computer for gravitational dynamics (Sugimoto et al. 1990),
we run N-body simulations of the dynamical evolution of 
stellar disks for  $16 - 20 t_{\rm dyn}$
($2.8 - 3.5$ Gyr) during which  stellar bars can be fully developed.
We investigate rotational velocities ($V_{\phi}$) at different $|z|$
and vertical metallicity gradients
for the bulge regions  at the final time step 
in models with different ${\alpha}_{\rm b, pds}$,
${\alpha}_{\rm b, tcds}$, ${\alpha}_{\rm v, pds}$ and ${\alpha}_{\rm v, tcds}$
both for PDS and TCDS. 
The total number of particles used for a model 
is 900000 in PDS and 1180000 in TCDS.
The adopted  gravitational softening
length is fixed at $0.014R_{\rm d}$, which corresponds to 252pc for
$R_{\rm d}=17.5$ kpc.

\begin{figure}
\psfig{file=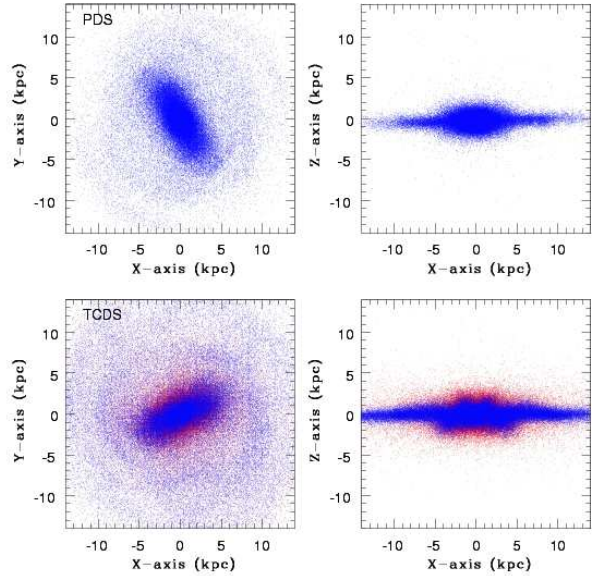,width=8.0cm}
\caption{
The final distributions of stars projected onto the $x$-$y$ (left) and
$x$-$z$ planes (right) in the model PDS1 (upper two) and those
in the model TCDS1 (lower two).  The blue and red particles represent stars
initially in the thin and thick disks, respectively, in the 
lower two panels.
}
\label{Figure. 1}
\end{figure}

\begin{figure}
\psfig{file=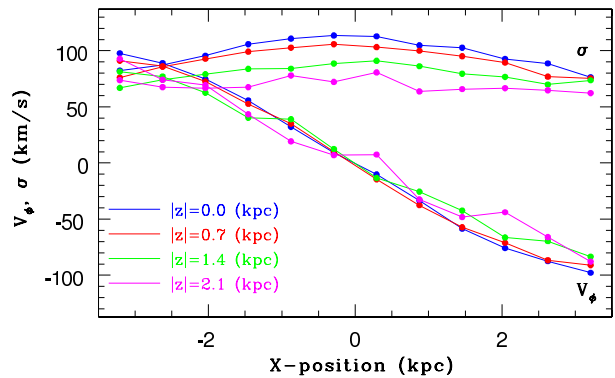,width=8.0cm}
\caption{
The rotational velocity ($V_{\phi}$) and velocity dispersion ($\sigma$)
along the $x$-axis of the simulated bulge (=bar) at different vertical
distances ($|z|$) in the model PDS1:
for $|z|=0$ kpc (blue), 0.7 kpc (red), 1.4 kpc (green), and 2.1 kpc (magenta).
}
\label{Figure. 2}
\end{figure}

\begin{figure}
\psfig{file=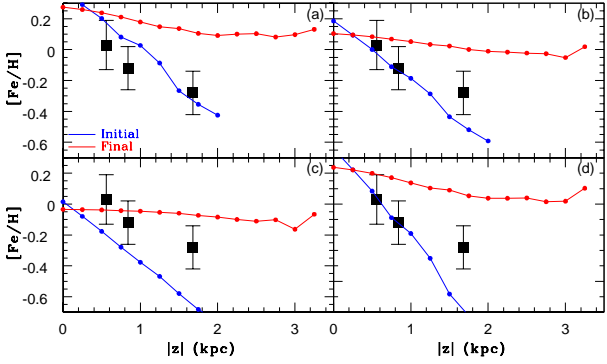,width=8.0cm}
\caption{
The initial (blue) and final (red) vertical metallicity gradients 
within the bulge region ($R<2$ kpc) in the model
PDS1 with different ${\alpha}_{\rm b, pds}$ and ${\alpha}_{\rm v, pds}$:
(a) ${\alpha}_{\rm b, pds}=-0.4$ and ${\alpha}_{\rm v, pds}=-0.4$,
(b) ${\alpha}_{\rm b, pds}=-0.2$ and ${\alpha}_{\rm v, pds}=-0.4$,
(c) ${\alpha}_{\rm b, pds}=0$ and ${\alpha}_{\rm v, pds}=-0.4$,
and (d) ${\alpha}_{\rm b, pds}=-0.4$ and ${\alpha}_{\rm v, pds}=-0.6$.
Three observational data points on mean metallicities
and dispersions at  $|z|=$ 0.56 kpc, 0.84 kpc, and 1.68 kpc
from Zoccali et al. (2008) are also shown  by filled squares and error bars,
respectively, in each frame for comparison.
}
\label{Figure. 3}
\end{figure}

\begin{figure}
\psfig{file=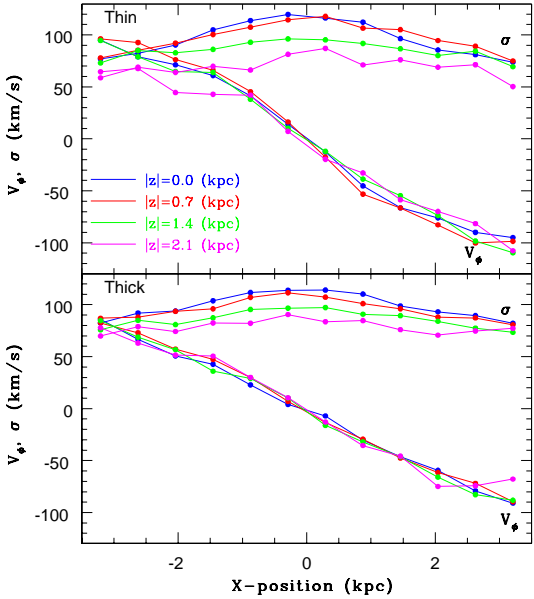,width=7.5cm}
\caption{
The same as Fig.2 but for the model TCDS1. Here $V_{\phi}$ and $\sigma$
profiles are shown separately for the thin disk (upper) and 
the thick one (lower)
for clarity.
}
\label{Figure. 4}
\end{figure}

\begin{figure}
\psfig{file=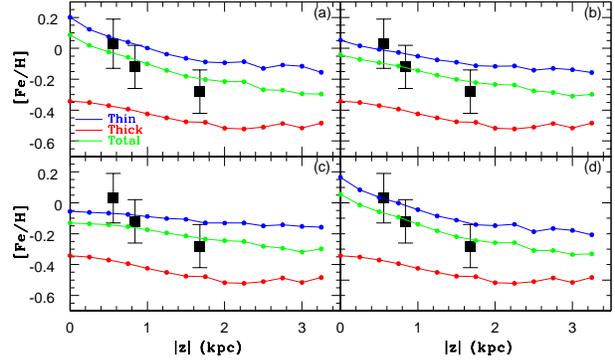,width=8.0cm}
\caption{
The same as Fig. 4 but for
the final vertical metallicity gradients for the thin disk (blue), 
thick one (red), and total (green) in the model TCDS 
with different ${\alpha}_{\rm b, tcds}$ and ${\alpha}_{\rm v, tcds}$:
(a) ${\alpha}_{\rm b, tcds}=-0.4$ and ${\alpha}_{\rm v, tcds}=-0.4$,
(b) ${\alpha}_{\rm b, tcds}=-0.2$ and ${\alpha}_{\rm v, tcds}=-0.4$,
(c) ${\alpha}_{\rm b, tcds}=0$ and ${\alpha}_{\rm v, tcds}=-0.4$,
and (d) ${\alpha}_{\rm b, tcds}=-0.4$ and ${\alpha}_{\rm v, tcds}=-0.6$.
}
\label{Figure. 5}
\end{figure}

\section{Results}

Fig. 1 shows that a strong  stellar bar is  fully developed 
by spontaneous bar/buckling  instability 
in the central region of the stellar disk  and the disk 
looks like a disk galaxy with a bulge in edge-on view
for the model PDS1.
As shown in Fig. 2, the central region of the disk 
with $R<3$ kpc 
clearly show rotation at different $|z|$ ($0 \le |z| \le 2.1$ kpc),
which demonstrates that the bulge has cylindrical rotation. 
The derived
amplitude ($\sim 100$ km s$^{-1}$) and profile of cylindrical rotation 
in this model are in good agreement  with the latest observational
results by Howard et al. (2009).
The velocity dispersion ($\sigma$) can be larger than $\sim 100$ km s$^{-1}$
in the center of the bar and 
depends on $|z|$ such that
$\sigma$ is larger for smaller $|z|$ in  the bulge.  
These results are also in good agreement 
with the observations by Howard et al. (2009).

Fig. 3 shows how  the initial vertical metallicity gradient of the stellar disk
in PDS
evolves  as the stellar bar change radial and vertical stellar structures owing
to its dynamical action on the disk.  
It is clear that irrespective of
${\alpha}_{\rm b, pds}$ and ${\alpha}_{\rm v, pds}$, 
the initial steep vertical gradient can become rather flattened
within 3.5 Gyr.
This is mainly because stellar populations that are located at different
$|z|$ and thus have different [Fe/H] can be mixed well during the bar instability
and the evolution of the bar. 
The simulated vertical metallicity gradients do not fit well with
the observational results, in particular,  for  $|z| \sim 1.7$ kpc,
which suggests that the bulge can not be simply formed from a pure thin 
one-component  stellar disk
through bar instability.

As shown in Fig. 1,  a stellar bar can be developed in the two-component
disk within $\sim 2$ Gyr in the model TCDS1
and it looks like a (boxy) bulge in edge-on view.
Although the thick disk  can not become a bar for  itself  owing
to the smaller mass fraction of the disk,  it can become a bar when
the thin disk  is transformed into a bar through bar instability. 
Fig. 4 clearly shows that both the thin and thick disks  show the maximum $V_{\phi}$
of $\sim 100$ km s$^{-1}$ and have cylindrical rotation with the amplitudes
and profiles similar to each other in the central 3 kpc.
The simulated maximum value of $\sigma$
and $\sigma$-profiles  dependent on $|z|$ in the thin and thick disks  are 
in good agreement 
with the observed ones. 
Although the thick disk  initially has higher velocity dispersions and a smaller amplitude
of rotation than the thin disk,  the two components can finally have similar kinematical
properties: the thick disk  has $\sigma$ at $|z|=1.4$kpc and 2.1 kpc  higher than 
those of the thin disk. 

%These  suggest that a significant fraction of stars in the present bulge 
%can originate from the inner part of the thick disk 
%and thus have a common origin with the thick disk stars.

Fig. 5 shows that although the initial vertical metallicity gradient of  
the bulge region in 
the thin disk  can be significantly flattened, 
the simulated profile for the whole  disk
can fit much better with the observational data in comparison with
those derived in the PDS1. The most important  reason for this
is that the mean metallicity at $|z|=1.7$ kpc in this model
can be as low as $-0.2$ owing to the presence of metal-poor stars 
of FD there: these metal-poor stars originate from
$R>3$ kpc of FD.
Since these stars  initially  stay at higher $|z|$ in the dynamically hotter thick disk,
they can not be strongly influenced by the bar 
and consequently they can stay longer at higher $|z|$ and thus can keep the lower
mean metallicity there. As a result of this, the mean metallicity of the whole
disk (bulge) at higher $|z|$ can keep lower.
In addition,
the bar in this model does not mix so well stellar populations 
with different
metallicities  in the thin disk
in comparison with the PDS1.
This less efficient mixing of stellar populations with different initial $|z|$
and [Fe/H] also contribute to the steeper vertical metallicity
gradient to some extent.
The results for three different models in Fig. 5
show  that steeper vertical metallicity gradients can be seen in models
with steeper  ${\alpha}_{\rm b, tcds}$ in TCDS.
Thus TCDS can reproduce reasonably well
the cylindrical rotation and vertical metallicity
gradient observed in  the Galactic bulge in a self-consistent manner.

\section{Discussion}

The results of the present numerical simulations suggest that
if the Galactic bulge was formed from an initially thin stellar disk
through bar instability, then the present bulge
is unlikely to have a steep vertical
metallicity gradient owing to the vertical mixing of stellar populations
with different metallicities by the bar.  However, it should be stressed
here that the present study is based on collisionless N-body simulations
that do not include gas dynamics which  could possibly suppress the formation
of strong  bars.
Accordingly it would  overestimate
the degree of  the vertical
mixing of stellar populations by stellar bars 
and 
thus can not completely rule out a possibility that dissipative gas
dynamics  can keep  the original vertical
metallicity gradient of the bulge.

In TCDS,  FD was formed with a star formation  time scale $t_{\rm sf}$ of $\sim 1 $ Gyr
about  10 Gyr ago through dissipative collapse
or merging of subgalactic clumps. 
Such a  short $t_{\rm sf}$  is suggested to be
consistent with  high [$\alpha$/Fe] observed in the bulge,
though $t_{\rm sf}$ could  be $\sim 0.1$ Gyr (e.g., Ballero et al. 2007).
Stars in the inner region of FS
finally become the old population of the bulge whereas those in the outer one become the thick
disk. 
After merger events about 10 Gyr ago, SD started to form 
and its inner part ($<2$kpc) could form  earlier and
much  more rapidly owing to its shorter dynamical
time scale in comparison with the outer part.  Stars in the inner part of SD can
finally become relatively younger populations 
of the bulge  compared with those in FD.
Owing to a rapid time scale of star formation in
the bulge composed of inner FD and SD stars, 
the chemical abundances of the old stellar populations in TCDS can be very
similar to what 
our previous bulge formation model with a rapid star formation
time scale predicted (Tsujimoto et al. 2010).
The model by Tsujimoto et al. (2010)
can explain the observed  MDF (Fulbright et al 2006) and
high [$\alpha$/Fe] 
%(McWilliam \& Rich 1994;
(Rich \& Origlia 2005; Fulbright et al. 2007; Johnson et al. 2011) for the bulge:
we thus suggest  that the present TCDS is in good agreement with 
these observations.
We will extensively investigate 
chemical abundances of the bulge
using a fully consistent
chemodynamical bulge model in our forthcoming papers.

The present TCDS suggests 
that the observed similarity in stellar 
populations between the bulge and the thick
disk (e.g., Mel\'endez et al. 2008; Bensby et al. 2010) can be naturally explained
in TCDS.
Also,  the simulated two-component bulge in the present study
can explain the observed presence of two distinct populations with
a varying mix of the two along the minor axis of the bulge
(Babusiaux et al. 2010). 
Thus the Galaxy might well have experienced merger events about 10 Gyr ago,
resulting in the transformation of a thin disk into 
either a thick disk or a flattened spheroid
supported mainly by rotation.

Previous chemical evolution models showed that the IMF needs to be significantly
flatter than the Salpeter one to explain the observed metallicity distribution
function (MDF) and higher [$\alpha$/Fe] of the bulge stars
(e.g., Tsujimoto et al. 2010).
On the other hand,  such a flat IMF is not necessary to explain the observed
chemical abundance of the thick disk stars around the solar neighborhood
(e.g.,  Chiappini et al. 1997).
In TCDS, the present bulge stars and thick disk ones around the solar neighborhood can
originate largely from inner and outer regions of FD, respectively.
Therefore, 
the IMF in FD
needs to depend on  $R$
such that it is flatter in the inner region
of the Galaxy.
This possible radial-dependent IMF of the Galaxy may well have
some important implications on the Galaxy evolution.

\section{Acknowledgment}
We are grateful to the  anonymous referee for valuable comments
which contribute to improve the present paper.
%KB acknowledge the financial support of the Australian Research Council
%throughout the course of this work.

\end{document}